\documentclass[
  journal=largetwo,
  manuscript=article-type,
  year=2020,
  volume=37,
]{cup-journal}

\usepackage{aas-macros}
\usepackage{amsmath, amssymb}
\usepackage[nopatch]{microtype}
\usepackage{booktabs}

\usepackage{multirow}
\usepackage{graphicx, float}	
\usepackage{subcaption} 
\usepackage{parcolumns} 

\usepackage{hyperref}
\hypersetup{
  colorlinks   = true, 
  urlcolor     = blue,
  linkcolor    = blue, 
  citecolor    = blue
}

\newcommand{\hoffmanninprep}{Hoffmann et al.\ in prep}
\newcommand{\bannisterinprep}{Bannister et al.\ in prep}
\newcommand{\wanginprep}{Wang et al.\ in prep}

\newcommand{\DMISM}{DM$_{\rm ISM}$}
\newcommand{\DMIGM}{DM$_{\rm cosmic}$}
\newcommand{\DMEG}{DM$_{\rm EG}$}
\newcommand{\DMG}{DM$_{\rm MW}$}
\newcommand{\DMhalo}{DM$_{\rm halo}$}
\newcommand{\DMobs}{DM$_{\rm obs}$}
\newcommand{\DMhost}{DM$_{\rm host}$}
\newcommand{\DMNE}{DM$_{\rm NE2001}$}
\newcommand{\DMunit}{pc\:cm$^{-3}$}

\newcommand{\DMmax}{DM$_{\rm max}$}

\newcommand{\sigmaISM}{$\sigma_{\rm ISM}$}
\newcommand{\muISM}{$\mu_{\rm ISM}$}
\newcommand{\sigmaHalo}{$\sigma_{\rm halo}$}
\newcommand{\muHalo}{$\mu_{\rm halo}$}

\newcommand{\nsfr}{$n$}
\newcommand{\Emax}{$E_{\mathrm{max}}$}
\newcommand{\Emin}{$E_{\mathrm{min}}$}

\newcommand{\FAST}{FAST}

\newcommand{\flyseye}{CRAFT Fly's Eye}
\newcommand{\icslow}{CRAFT/ICS 900 MHz}
\newcommand{\icsmid}{CRAFT/ICS 1.3 GHz}
\newcommand{\icshigh}{CRAFT/ICS 1.6 GHz}
\newcommand{\parkes}{Parkes/Mb}

\newcommand{\myedit}[1]{\textcolor{black}{#1}}

\title{Modelling DSA, FAST and CRAFT surveys in a z-DM analysis and constraining a minimum FRB energy}

\author{J.~Hoffmann}
\affiliation{International Centre for Radio Astronomy Research, Curtin University, Bentley, WA 6102, Australia}
\email[J. Hoffmann]{jordan.hoffmann@postgrad.curtin.edu.au}

\author{C.~W.~James}
\affiliation{International Centre for Radio Astronomy Research, Curtin University, Bentley, WA 6102, Australia}

\author{M.~Glowacki}
\affiliation{International Centre for Radio Astronomy Research, Curtin University, Bentley, WA 6102, Australia}

\author{J.~X.~Prochaska}
\affiliation{Department of Astronomy and Astrophysics, University of California, Santa Cruz, CA 95064, USA}
\alsoaffiliation{Kavli Institute for the Physics and Mathematics of the Universe, 5-1-5 Kashiwanoha, Kashiwa 277-8583, Japan}
\alsoaffiliation{Division of Science, National Astronomical Observatory of Japan, 2-21-1 Osawa, Mitaka, Tokyo 181-8588, Japan}

\author{A.~C.~Gordon}
\affiliation{Center for Interdisciplinary Exploration and Research in Astrophysics (CIERA) and Department of Physics and Astronomy, Northwestern University, Evanston, IL 60208, USA}

\author{A.~T.~Deller}
\affiliation{Centre for Astrophysics and Supercomputing, Swinburne University of Technology, Hawthorn, VIC, 3122, Australia}

\author{R.~M.~Shannon}
\affiliation{Centre for Astrophysics and Supercomputing, Swinburne University of Technology, Hawthorn, VIC, 3122, Australia}

\author{S.~D.~Ryder}
\affiliation{School of Mathematical and Physical Sciences, Macquarie University, NSW 2109, Australia}
\alsoaffiliation{Astrophysics and Space Technologies Research Centre, Macquarie University, Sydney, NSW 2109, Australia}

\keywords{cosmological parameters} 

\begin{document}

\begin{abstract}
Fast radio burst (FRB) science primarily revolves around two facets: the origin of these bursts and their use in cosmological studies. This work follows from previous redshift--dispersion measure ($z$--DM) analyses in which we model instrumental biases and simultaneously fit population parameters and cosmological parameters to the observed population of FRBs. This sheds light on both the progenitors of FRBs and cosmological questions. Previously, we have completed similar analyses with data from the Australian Square Kilometer Array Pathfinder (ASKAP) and the Murriyang (Parkes) Multibeam system. \myedit{In this manuscript, we use 119 FRBs with 29 associated redshifts by additionally modelling the Deep Synoptic Array (DSA) and the Five-hundred-meter Aperture Spherical radio Telescope (FAST).} We also invoke a Markov chain Monte Carlo (MCMC) sampler and implement uncertainty in the Galactic DM contributions. The latter leads to larger uncertainties in derived model parameters than previous estimates despite the additional data \myedit{and indicate that precise measurements of \DMISM{} will be important in the future}. We provide refined constraints on FRB population parameters and derive a new constraint on the minimum FRB energy of \myedit{log\,\Emin{}(erg)=39.47$^{+0.54}_{-1.28}$} which is significantly higher than bursts detected from strong repeaters. This result likely indicates a low-energy turnover in the luminosity function or may alternatively suggest that strong repeaters have a different luminosity function to single bursts. We also predict that FAST will detect 25 -- 41\% of their FRBs at $z \gtrsim 2$ and DSA will detect 2 -- 12\% of their FRBs at $z \gtrsim 1$.
\end{abstract}

\section{Introduction} \label{sec:Introduction}
Fast radio bursts (FRBs) are highly energetic extragalactic bursts of radio waves lasting of order milliseconds in duration \citep[e.g][]{Lorimer2007, Thornton2013}. Since their discovery, many progenitor models have been suggested \citep{platts2019}, however, a definitive model has not yet arisen although magnetar origins are favoured within the community. 



The population parameters of FRBs can inform one about their progenitor objects and emission mechanisms. The luminosity function and spectral dependence of FRBs can give insights into the underlying physical processes as differing emission mechanisms have unique energetic and spectral behaviours \citep{Lu2019, Luo2020, Macquart2020, Arcus2020, james2022, shin2023}. Similarly, the evolution of FRB progenitor objects through cosmic time also constrains their origin as many predicted progenitors have expected cosmological evolution models \citep{Macquart2018b, Zhang2021}. As such, FRB population parameters offer a powerful means to determine the progenitors of FRBs \citep{Luo2018, james2022, shin2023}.


As radiation passes through cool plasmas, it slows down in a predictable frequency-dependent manner. This can be observationally determined as the dispersion measure (DM) which is a direct quantification of the integrated column density of electrons along the line of sight. A combination of the DM and host galaxy redshift, $z$, then enables studies of the cosmological electron distribution which traces the cosmological baryon distribution in an ionised universe. Thus, FRBs have found immense success in cosmological studies. \citet{Macquart2020} used FRBs to find the `missing baryons' \citep{fukugita1998} in the Universe and \citet{james2022b} demonstrated that FRBs could be used to shed light on the Hubble constant tension given a sufficient number of localised FRBs. The most recent efforts examine the distribution of cosmic baryons through FRB population studies \citep{Baptista2023} and cross-correlation analyses \citep{Khrykin2024}.

This work builds upon the work of \citet{james2022b}, \citet{Baptista2023} and the \verb|zDM| code used therein \citep{zdm}. The \verb|zDM| code was developed to model the $z$--DM relation for FRBs detected with the Murriyang (Parkes) Multibeam system \citep[\parkes{};][]{Hobbs2020} and the Australian Square Kilometre Array Pathfinder \citep[ASKAP;][]{Hotan2021} under the Commensal Real-time ASKAP Fast Transients (CRAFT) survey. It accounts for telescope biases and fits FRB population parameters and cosmological constants. The most recent analysis of \citet{Baptista2023} used 78 FRBs of which 21 were localised but was still limited by small-number statistics. Hence, the most effective way to obtain more stringent constraints is to include more FRBs, and currently the most efficient approach is to include additional FRB surveys. In addition to Parkes and ASKAP, the Deep Synoptic Array \citep[DSA;][]{dsa102019}, the Five-hundred-meter Aperture Spherical radio Telescope \citep[FAST;][]{FAST, FAST2}, the Canadian Hydrogen Intensity Mapping Experiment \citep[CHIME;][]{CHIME}, MeerKAT \citep{jonas2016} and the Upgraded Molongolo Observatory Synthesis Telescope \citep[UTMOST;][]{UTMOST} have discovered significant numbers of FRBs in blind searches. However, to include these FRBs in an unbiased way, we must model the instrumental biases of each telescope to ensure the same intrinsic population is being sampled.

In this work, we focus on the modelling and inclusion of the FAST and DSA surveys. While FAST has a relatively small number of FRBs detected in the two blind surveys conducted \citep{Zhu2020, Niu2021, Zhou2023} -- of which none have been localised -- its high sensitivity allows it to probe a new region of the parameter space and hence is of interest. Conversely, DSA probes a similar parameter space to ASKAP but has a large number of FRBs and localisations. The telescope is expected to continue to detect many FRBs and associate them with host galaxies and therefore lends valuable information to a $z$-DM analysis. The telescope is also located in the Northern Hemisphere and hence has a sky coverage that is complementary to that of ASKAP allowing for a more uniform sampling of the sky.

In Section \ref{sec:surveys} we describe our models for FAST and DSA alongside all of the relevant telescope and search algorithm parameters. We also outline additional CRAFT FRBs that were not used in previous analyses. We then describe improvements that have been made to the analysis in Section \ref{sec:improvements}. We give new parameter constraints in Section \ref{sec:results} and give our predictions for the $z$ and DM distributions of FAST and DSA in Section \ref{sec:predictions}. Finally, we outline our conclusions in Section \ref{sec:conclusions}.

\section{Survey description and data} \label{sec:surveys}
In addition to the FRBs used in the analysis of \citet{james2022b} and \citet{Baptista2023}, we include more recent FRB discoveries of CRAFT as well as those of FAST and DSA in this work.

\subsection{Exclusion of large surveys} \label{sec:exclusion}
We choose not to include CHIME, MeerKAT or UTMOST in this analysis, even though they do have a significant number of FRBs detected in their blind searches, for the following reasons. 

\begin{itemize}
    \item CHIME has a unique observational bias towards detecting repeating sources in comparison to the other instruments. \citet{James2023} has shown that a $z$-DM analysis with CHIME is not meaningful unless repeaters are additionally modelled. As such, we leave the inclusion of this instrument to future work (\hoffmanninprep). \citet{shin2023} previously conducted a population parameter analysis for CHIME which yielded similar results to \citet{james2022b} and hence we expect the inclusion of this survey to give stronger constraints on the parameters without significantly changing our conclusions.

    \item MeerKAT detects FRBs in three different modes and does not yet have a catalogue published. Notable events may thus be published first, potentially resulting in a reporting bias. We therefore do not consider MeerKAT FRBs either, however, once a more complete catalogue is published we believe that MeerKAT will contribute significantly given its sensitivity. 

    \item UTMOST has a complex beam pattern and is thus difficult to model accurately. Additionally, none of these FRBs have an associated $z$ and hence we do not believe the inclusion of this survey is currently worth the difficulty of modelling the telescope. 
\end{itemize}


\subsection{Modelling FAST} \label{sec:FAST}
FAST is a spherical single-dish telescope located in China. The spherical reflector has a curvature following a radius of 300\,m, an aperture diameter of 500\,m and an illuminated aperture diameter of 300\,m. This large dish size results in FAST having a high sensitivity but a small field of view. Hence, it can probe deeper into the Universe and thus explore a new region of the parameter space in comparison to less sensitive wide-field survey instruments such as ASKAP. To date, FAST has published nine new FRBs \citep{Zhu2020, Niu2021, Zhou2023}. These FRBs have DMs ranging from 1187.7 to 2765.2\,\DMunit{} with an average of $\sim$1800\,\DMunit{} (although for half of the sample $\sim 500$\,\DMunit{} is from \DMISM{}) which greatly exceeds that of any other survey. However, FAST does not have sufficient resolution to consistently associate FRBs with a host galaxy, particularly at higher redshifts where we expect these FRBs to come from. While FRBs which have an associated redshift hold the greatest constraining power in a $z$--DM analysis, the fact that these FRBs lie in a unique portion of the parameter space still makes them useful. If they are detected to repeat, follow-up can be conducted with other instruments that could enable an association with a host galaxy which is an exciting prospect. 

To include FAST FRBs in our analysis, we must have a beam model to account for telescope biases. The current FAST receiver consists of an array of 19 beams operating at frequencies of 1.05 to 1.45\,GHz. The beams have a hexagonal layout and are spaced $\sim$5$^{\prime}$ apart \citep[exact spacings are given in][]{Jiang2019J}. The half-power beamwidth (HPBW) for each beam varies from $\sim$2.8$^{\prime}$ to $\sim$3.5$^{\prime}$ over the band, with each beam varying marginally around these values. We model the beam pattern by treating each of the 19 beams as Gaussian beams with HPBW, location, and sensitivity given by the values at 1250\,MHz from \citet{Jiang2019J}. We then superimpose the 19 beams in the relevant configuration and coarsely discretise the beam solid angle into 10 bins of sensitivity due to computational limitations.

In Table~\ref{table:FASTparams}, we present other parameters of FAST and the associated FRB searches. The FRBs used in this analysis are presented in Table~\ref{table:FAST}.


\begin{table}
\begin{center}
\caption{Relevant parameters of the FAST telescope and FRB searches to a $z$--DM analysis. The values presented are taken or derived from \citet{Zhu2020}, \citet{Niu2021} and \citet{Zhou2023}.}
\label{table:FASTparams}
\begin{tabular}{lcc}
\hline
Parameter & Value \\
\hline 
Central frequency (MHz) & 1250 \\
Bandwidth (MHz) & 500 \\
Channel width (kHz) & 122 \\
Time resolution ($\mu$s) & 196.608 \\
SNR threshold & 7.0 \\
Fluence threshold (Jy\,ms) & 0.0146 \\
\hline
\end{tabular}
\end{center}
\end{table}

\begin{table}
\begin{center}
\caption{FAST FRBs used in this analysis. Given is the internal name, observed DM, DM contribution from the ISM estimated by the NE2001 model \citep{Cordes2002} and SNR at detection.}
\label{table:FAST}
\begin{tabular}{cccccl}
\hline 
Name & \DMobs{} & \DMNE{} & SNR & Ref.\ \\
& (\DMunit{}) & (\DMunit{}) & &
\\ 
\hline 
181123 & 1812.0 & 97.17 & 95 & \citet{Zhu2020} \\ 
\hline
181017 & 1845.2 & 34.96 & 17 & \\ 
181118 & 1187.7 & 69.57 & 13 & \citet{Niu2021} \\ 
181130 & 1705.5 & 38.57 & 26 & \\ 
\hline 
210126 & 1990.4 & 724 & 13.60 & \\
210208 & 1448.4 & 432 & 52.54 & \\
210705 & 2011.6 & 484 & 23.82 & \citet{Zhou2023} \\
211005 & 2765.2 & 528 & 20.06 & \\
220306 & 1273.9 & 454 & 20.98 & \\
\hline
\end{tabular} 
\end{center} 
\end{table}

\subsection{Modelling DSA} \label{sec:DSA}
We consider 25 new FRBs detected by DSA-110 during its commissioning observations \citep{Sherman2023}. The array consists of 48 core antennas along an east-west line which are densely packed with a maximum spacing of 400\,m \citep{Ravi2023b}. An additional 15 outrigger antennas are used in the localisation of the sources after detection in an attempt to identify the host galaxies of the FRBs to subsequently enable redshifts to be obtained. Each antenna has a diameter of 4.65\,m, a typical system temperature of 25\,K and observes at 1405\,MHz with a 187.5\,MHz bandwidth. Data from each of the 48 core antennas are coherently combined to form 256 fan-shaped search beams separated by 1$^{\prime}$ to span 4.27$^{\circ}$. The exact spacing of the antennas has not yet been made publicly available so we cannot precisely model the beam pattern. Regardless, the spacing is dense enough that these search beams have significant overlap with each other. Furthermore, were the antennas to be equally distributed over the 400\,m east-west line, their grating lobe spacing at zenith would be 1.47$^{\circ}$, suggesting significant sensitivity outside the nominal 4.27$^{\circ}$ spanned by the formed beams. Hence, we approximate the formed beam pattern by the primary beam pattern. We model the beam shape of DSA-110 during this commissioning phase as a Gaussian with a full-width half-maximum (FWHM) of 2.6$^{\circ}$, corresponding to the 4.65 m dish size. We note that \citet{Ravi2023b} instead quote the FWHM as 3.4$^{\circ}$. However, the total Gaussian width --- as opposed to the shape --- only affects the total number of FRBs detected which we do not consider, as the total effective observation time, $T_{\mathrm{obs}}$, is not known. Hence, this discrepancy has no impact on our results. Similarly to FAST modelling, we then discretise the beam into 10 bins due to computational limitations.

We present other parameters of the DSA-110 commissioning observations that are relevant to a $z$--DM analysis in Table~\ref{table:DSAparams}. The values presented are either taken directly from \citet{Ravi2023b} and \citet{Sherman2023} or derived from values therein. 

The relevant properties of each DSA FRB are presented in Table~\ref{table:DSA} and are taken from \citet{Sherman2023} and \citet{Law2023}. Redshifts for 12 of the 25 DSA FRBs are presented in \citet{Law2023}; however, it is not possible to use all 12 redshifts in an unbiased way. Because higher redshift FRBs will, on average, be hosted by apparently fainter galaxies, an incomplete sample is likely biased against high-$z$ FRBs. In turn, this biases the sample {\it towards} lower redshifts at a given DM. To avoid such a bias, we only utilise $z$ values for FRBs below a maximum \DMEG{} value (see Section \ref{sec:uDMG} for definitions of each DM contribution). This maximum value corresponds to the limit for which all FRBs below this threshold have an associated $z$ and hence we have no concerns regarding not detecting high-$z$ host galaxies. When considering such a limit, we do not consider uncertainty in \DMG{}. As such, \DMhalo{} is constant over each of the FRBs and hence placing a cutoff on \DMEG{} is equivalent to placing a cutoff on \DMobs{}$-$\DMISM{}. For this survey, we find that this limit is at \DMobs{}$-$\DMISM{}$= 183$\,\DMunit{}. Thus, only FRBs 20220207C, 20220319D and 202205509G have redshifts that can be utilised in an unbiased way. For the rest, we use the probability of \DMEG{}, $P($\DMEG{}), in place of $P(z$, \DMEG{}).

\begin{table}
\begin{center}
\caption{Relevant parameters of DSA-110 commissioning observations to a $z$-DM analysis. The values presented are taken or derived from \citet{Ravi2023b} and \citet{Sherman2023}.}
\label{table:DSAparams}
\begin{tabular}{lc}
\hline
Parameter & Value \\
\hline 
Central frequency (MHz) & 1405 \\
Bandwidth (MHz) & 187.5 \\
Channel width (kHz) & 244.141 \\
Time resolution ($\mu$s) & 262.144 \\
SNR threshold & 8.5 \\
Fluence threshold (Jy\,ms) & 1.96 \\
\hline
\end{tabular}
\end{center}
\end{table}

\begin{table*}
\begin{center}
\caption{DSA FRBs used in this analysis. Given is the TNS name, observed DM, DM contribution from the ISM estimated by the NE2001 model \citep{Cordes2002}, observed $z$, probability of association with the identified host galaxy and whether or not the localisation was used in this analysis. We do not utilise the redshifts of some FRBs to avoid a detection bias against FRBs with a high $z$ for their DM. Bolded rows show the FRBs where we use $z$ information. All FRBs are presented in \citet{Sherman2023} and localisations are presented in \citet{Law2023}.}
\label{table:DSA}
\begin{tabular}{cccccccc}
\hline 
Name & \DMobs{} & \DMNE{} & SNR & $z$ & $P_{\mathrm{host}}$ & Used $z$ \\
& (\DMunit{}) & (\DMunit{}) & & & & (Y/N) \\ 
\hline 
20220121B & 313.421 & 79.99 & 9.38 & - & - & - \\
20220204A & 612.584 & 52.58 & 16.22 & - & - & - \\
\textbf{20220207C} & \textbf{263.0} & \textbf{74.99} & \textbf{59.96} & \textbf{0.043} & \textbf{0.99} & \textbf{Y} \\
20220208A & 440.73 & 88.37 & 13.77 & - & - & - \\
20220307B & 499.328 & 120.02 & 11.91 & 0.248 & 0.99 & N \\
20220310F & 462.657 & 45.46 & 68.41 & 0.478 & 0.99 & N \\
\textbf{20220319D} & \textbf{110.95} & \textbf{126.77} & \textbf{79.0} & \textbf{0.011} & \textbf{0.99} & \textbf{Y} \\
20220330D & 467.788 & 38.42 & 12.94 & - & - & - \\
20220418A & 624.124 & 36.35 & 10.88 & 0.622 & 0.97 & N \\
20220424E & 863.932 & 132.80 & 9.41 & - & - & - \\
20220506D & 396.651 & 82.85 & 48.92 & 0.300 & 0.98 & N \\
\textbf{20220509G} & \textbf{270.26} & \textbf{55.28} & \textbf{21.51} & \textbf{0.089} & \textbf{0.99} & \textbf{Y} \\
20220726A & 686.232 & 79.72 & 12.72 & - & - & - \\
20220801A & 413.416 & 101.63 & 9.25 & - & - & - \\
20220825A & 649.893 & 77.31 & 15.06 & 0.241 & 1.0 & N \\
20220831A & 1146.14 & 105.95 & 19.19 & - & - & - \\
20220914A & 630.703 & 54.39 & 9.64 & 0.114 & 0.97 & N \\
20220920A & 314.977 & 39.64 & 14.35 & 0.158 & 0.99 & N \\
20220926A & 441.984 & 104.28 & 10.26 & - & - & - \\
20221002A & 319.951 & 51.47 & 8.50 & - & - & - \\
20221012A & 440.358 & 54.06 & 9.41 & 0.285 & 1.0 & N \\
20221027A & 452.723 & 47.13 & 12.13 & - & - & - \\
20221029A & 1391.746 & 43.13 & 12.06 & - & - & - \\
20221101B & 491.554 & 116.47 & 10.12 & - & - & - \\
20221101A & 1475.53 & 79.69 & 14.97 & - & - & - \\
\hline
\end{tabular} 
\end{center} 
\end{table*} 

\begin{table*}
\begin{center}
\caption{Additional CRAFT FRBs used in this analysis that were not included in the analysis of \citet{james2022b} or \citet{Baptista2023}. Given is the TNS name, observed DM, DM contribution from the ISM estimated by the NE2001 model \citep{Cordes2002}, central observational frequency and observed $z$. All FRBs presented here are from \citet{ICS2024}. A channel width of 1\,MHz and a time resolution of 1.182\,ms were utilised during the searches. We assume an SNR threshold of 14 and hence do not list FRBs below this threshold. In actual searches, a threshold of 9 was used.}
\label{table:CRAFT}
\begin{tabular}{cccccc}
\hline 
Name & \DMobs{} & \DMISM{} & $\nu$ & SNR & $z$ \\ 
& (\DMunit{}) & (\DMunit{}) & MHz & & \\ 
\hline 
\multicolumn{6}{c}{\icslow{}} \\
\hline
20230521A & 640.2 & 41.8 & 831.5 & 15.2 & - \\
20230708A & 411.5 & 50.2 & 920.5 & 31.5 & 0.105 \\
20230902A & 440.1 & 34.3 & 831.5 & 11.8 & - \\
20231006A & 509.7 & 67.5 & 863.5 & 15.2 & - \\
20231226A & 329.9 & 38.0 & 863.5 & 17.8 & - \\
\hline
\multicolumn{6}{c}{\icsmid{}} \\
\hline
20230526A & 316.4 & 50.0 & 1271.5 & 22.1 & 0.157 \\
20230718A & 477.0 & 395.6 & 1271.5 & 10.9 & 0.0358 \\
20230731A & 701.1 & 547.1 & 1271.5 & 16.6 & - \\
\hline
\end{tabular} 
\end{center} 
\end{table*} 

\subsection{Updated CRAFT surveys} \label{sec:CRAFT}
In the analysis of \citet{james2022b} and \citet{Baptista2023}, three broad samples of FRBs are used. That is, FRBs detected by the Murriyang (Parkes) Multibeam system \citep[\parkes{}; e.g.][]{StaveleySmith1996, Keane2018}; FRBs detected by ASKAP in the Fly's Eye mode \citep[\flyseye{};][]{Bannister2017}; and FRBs detected by ASKAP in the incoherent sum mode \citep[CRAFT/ICS][]{Bannister2019b, ICS2024}. The CRAFT/ICS FRBs are further divided into three frequency categories (\icslow{}, \icsmid{} and \icshigh{}) within which we approximate all of the central observational frequencies by the average value in that category. In addition to the FRBs used in the previous analysis, we also include the FRBs listed in Table~\ref{table:CRAFT}, which are reported in \citet{ICS2024} and include all FRBs detected by CRAFT up until the end of 2023. For these additional FRBs, a channel width of 1\,MHz, a time resolution of 1.182\,ms and an SNR threshold of 9 were utilised during the searches. We also choose to exclude FRB 20171216A from the Fly's Eye survey which was previously included as it has a reported SNR of 8.0 which is below the SNR threshold of 9.5.


\section{Modifications to the analysis} \label{sec:improvements}
This work is an extension of \citet{james2022b} and hence we consider the same parameters. That is, the correlation between the abundance of FRB progenitors and the star formation rate (SFR) history of the Universe, \nsfr{}; the frequency dependence of the FRB event rate, $\alpha$; the mean ($\mu_{\mathrm{host}}$) and standard deviation ($\sigma_{\mathrm{host}}$) of the log-normally distributed \DMhost{} contribution; the turnover of the luminosity function when modelled as a Gamma function, \Emax{}; the integrated slope of the luminosity function, $\gamma$; and the Hubble constant, $H_0$. We additionally include a hard cut-off in the luminosity function at some minimum energy, \Emin{}, as a free parameter as discussed in Section \ref{sec:Emin}. 

We adopt the same general methodology and models as those described in \citet{james2022b} and here we discuss improvements and adjustments to those methods.

\subsection{Incorporating uncertainty in \DMG{}} \label{sec:uDMG}
Our analysis considers two components of the Galactic contributions to DM; namely contributions from the plasma in the interstellar medium (\DMISM{}) and from the baryonic matter embedded in the Milky Way's dark-matter halo (\DMhalo{}). In previous studies, we used the NE2001 model \citep{Cordes2002} to estimate \DMISM{} and did not consider any uncertainties. We additionally assumed a constant \DMhalo{} of 50 \DMunit{} which assumes an isotropic halo and ignores fluctuations between differing lines of sight. By not considering uncertainties in both of these parameters, we naturally overestimate the precision of the resulting parameter constraints. Furthermore, FRBs that have a low \DMIGM{} value for their corresponding $z$ give strong constraints on the cosmic baryon density of the Universe and hence hold a large amount of the constraining power for cosmological constants. Therefore, overestimating \DMG{} (which underestimates \DMIGM{}) for these FRBs can artificially make their constraining power more significant and hence can skew the resulting analysis.

A more extreme case of this is seen in FRB 20220319D. This FRB was recently detected by DSA with an estimated \DMISM{} that exceeds the total DM of the FRB \citep{Ravi2023}. Localisation of the FRB to a host galaxy with a high likelihood suggests that the burst is extragalactic which mandates \DMobs{} > \DMG{}{}. Hence, having \DMISM{} > \DMobs{} is an unphysical scenario. Therefore, to include such an FRB in our $z$--DM analysis, we must consider uncertainties from measurement error and/or physical scatter in \DMG{}.


To quantify the uncertainty in \DMG{}, we consider distributions of \DMISM{} and \DMhalo{} individually. The two prevailing models for estimating \DMISM{} contributions are the NE2001 \citep{Cordes2002} and YMW16 \citep{Yao2017} models which make use of observed pulsar DMs. While such models are the best available, they are known to be unreliable and are often inconsistent with each other \citep[e.g][]{pygedm}. The exact uncertainties for these values are unclear, however, these estimates are typically accurate to a factor of two \citep{Schnitzeler2012}. Thus, for each FRB, we model $P$(\DMISM{} $\mid$ \DMobs{}, \DMNE{}) as a normal distribution with mean \muISM{}=\DMNE{} and standard deviation \sigmaISM{}=\DMNE{}/2. Additionally, we must satisfy 0 \DMunit{} < \DMISM{} < \DMobs{} for a physical scenario and hence we truncate the distribution at these limits.

The MW halo is expected to be approximately spherical; however, the mean is uncertain and directionally dependent fluctuations are possible. \citet{Prochaska2019a} suggest a nominal mean \DMhalo{} contribution of 50--80\,\DMunit{} and we continue to estimate the mean as 50\,\DMunit{} due to the presence of low DM FRBs such as FRB 202200319D. We additionally implement an uncertainty in \DMhalo{} which we hope also accounts for fluctuations along different lines of sight. As such, we model $P$(\DMhalo{} $\mid$ \DMobs{}, \DMNE{}) as a normal distribution with a mean of \muHalo{}~=~50\,\DMunit{}, and a standard deviation of \sigmaHalo{}~=~15\,\DMunit{}. Similarly to \DMISM{}, we truncate this distribution at the physical limits of 0 \DMunit{} and \DMobs{}.

The total Galactic contribution is simply the sum of \DMISM{} and \DMhalo{} and hence we determine the distribution of $P($\DMG{} $\mid$ \DMobs{}, \DMNE{}) by taking the convolution of our distributions of $P$(\DMISM{} $\mid$ \DMobs{}, \DMNE{}) and $P$(\DMhalo{} $\mid$ \DMobs{}, \DMNE{}). We numerically calculate the convolution as the distributions are truncated Gaussians and hence are not easily determined analytically. We then truncate this distribution again at the physical limits of 0 \DMunit{} and \DMobs{}. We note that we do not renormalise any of the distributions after truncation as doing so discards the probability that this FRB was detected at all. 

\DMEG{} is given by
\begin{equation}
    \mathrm{DM}_\mathrm{EG} = \mathrm{DM_{obs}} - \mathrm{DM_{MW}},
\end{equation}
and hence this gives a distribution of $P($\DMEG{} $\mid$ \DMobs{}, \DMNE{}). The probability of detecting an FRB at the observed DM is then
\begin{eqnarray}
    P(\mathrm{DM_{obs}}) = \int_{0}^{\mathrm{DM_{obs}}} P(\mathrm{DM_{EG}} \mid \mathrm{DM_{obs}}, \mathrm{DM_{NE2001}}) \nonumber \\
    \times P(\mathrm{DM_{EG}} \mid \pmb{\theta}) \: d\mathrm{DM_{EG}},
\end{eqnarray}
where $\pmb{\theta}$ represents a vector of all the model parameters and hence $P($\DMEG{}\,$\mid \pmb{\theta})$ is the probability of \DMEG{} given a set of model parameters and the detection biases of the survey. Previously, we considered $P($\DMEG{}\,$\mid$\,\DMobs{},\,\DMNE{}) to be a delta function at \DMobs{} - \DMNE{} and hence only considered  $P($\DMEG{}\,$\mid \,\pmb{\theta})$.

\subsection{Search limits in DM}

We currently use the analytical approximation of \citet{Cordes2003} to estimate the DM-dependent sensitivity of FRB searches. While pulse injection characterises this sensitivity more accurately, these deviations make minimal differences \citep{Qiu2023, Hoffmann2024}. 

FRB searches are computationally limited and hence implement a maximum DM to which searches are conducted, \DMmax{}. As \DMobs{} must be less than \DMmax{} for a detection to occur (excluding the rare event of extremely bright FRBs above the maximum searched DM), $P($\DMobs{}) is not dependent on this search limit. Therefore, the only impact from \DMmax{} not being considered is in $P(N)$ -- the expected number of events for each survey. Hence, when determining $P(N)$ we now only consider DMs up to \DMmax{}.

For CRAFT surveys, the maximum DM used corresponds to 4096 time samples. We do not consider \DMmax{} for the \parkes{} survey. DSA uses a maximum search DM of 1500\,\DMunit{} \citep{Law2023} and FAST uses a maximum DM of 5000\,\DMunit{} in the Commensal Radio Astronomy FAST Survey \citep[CRAFTS;][]{Niu2021, CRAFTS} and 3700\,\DMunit{} in the Galactic Plane Pulsar Snapshot \citep[GPPS;][]{Zhou2023} survey. We model both surveys as a single survey for computational ease as the maximum searched DM is the only difference and we do not expect this to have any significant contributions. We approximate the maximum searched DM as 4350\,\DMunit{}.

\subsection{MCMC implementation} \label{sec:mcmc}
The \verb|zDM| code was implemented to calculate likelihoods over a cube of the parameters. While such an implementation was possible at that time, the introduction of additional surveys and free parameters increases the computational load by orders of magnitude and hence makes such an analysis impractical. As such, we use a Python implementation of a Markov-Chain Monte-Carlo (MCMC) sampler, \textsc{emcee}, which allows the code to be scaled to incorporate additional parameters without significant additional computational cost \citep{emcee}. Each likelihood calculation for a single set of model parameters typically takes of order $\sim$75\,s. Hence, we use a parallelised version which we run on the OzSTAR supercomputer based at Swinburne University of Technology. We use 30 walkers each running for 2400 steps with a burn-in of 300 samples.

\subsection{Allowing \Emin{} as a free parameter} \label{sec:Emin}
The intrinsic luminosity function of FRBs is not well known. As such, we model it as a Gamma function to avoid having a sharp cutoff at high energies which is otherwise implemented in a simple power law model. However, we still use a sharp cutoff for the minimum burst energy \Emin{}. This minimum burst energy has an analogous effect to the SNR threshold for surveys as it restricts the detection of low-fluence FRBs. The only difference is that \Emin{} introduces a $z$ dependence to this threshold and hence is more constraining for nearby FRBs. In the analysis of \citet{james2022b} we set \Emin{} to a conservatively low value of $10^{30}$\,ergs which is well below the corresponding SNR thresholds for the considered surveys. This value is low enough that the SNR threshold is the primary limit on low-energy detections and hence makes the assumption that these surveys are not sensitive enough to probe \Emin{}.

In this work, we include FAST which is significantly more sensitive than the Parkes and ASKAP radio telescopes. As such, it is more likely to be able to probe \Emin{}. With the addition of the MCMC sampler, allowing additional parameters to vary does not significantly increase the computational load. We therefore allow \Emin{} to vary as a free parameter in this work.

\section{Parameter constraints} \label{sec:results}
For the following analyses, we take a uniform prior for each parameter as described in Table~\ref{table:params}. For parameters that are well constrained within the priors, the chosen limits are arbitrary (i.e. all except \Emin{}, \Emax{} and $H_0$). We broadly note that \Emin{} and \Emax{} have large tails in all cases and hence the quoted values do not directly portray the 16\%, 50\% and 84\% quantiles as for the other parameters.

\begin{figure*}
\begin{center}
\includegraphics[width=\linewidth]{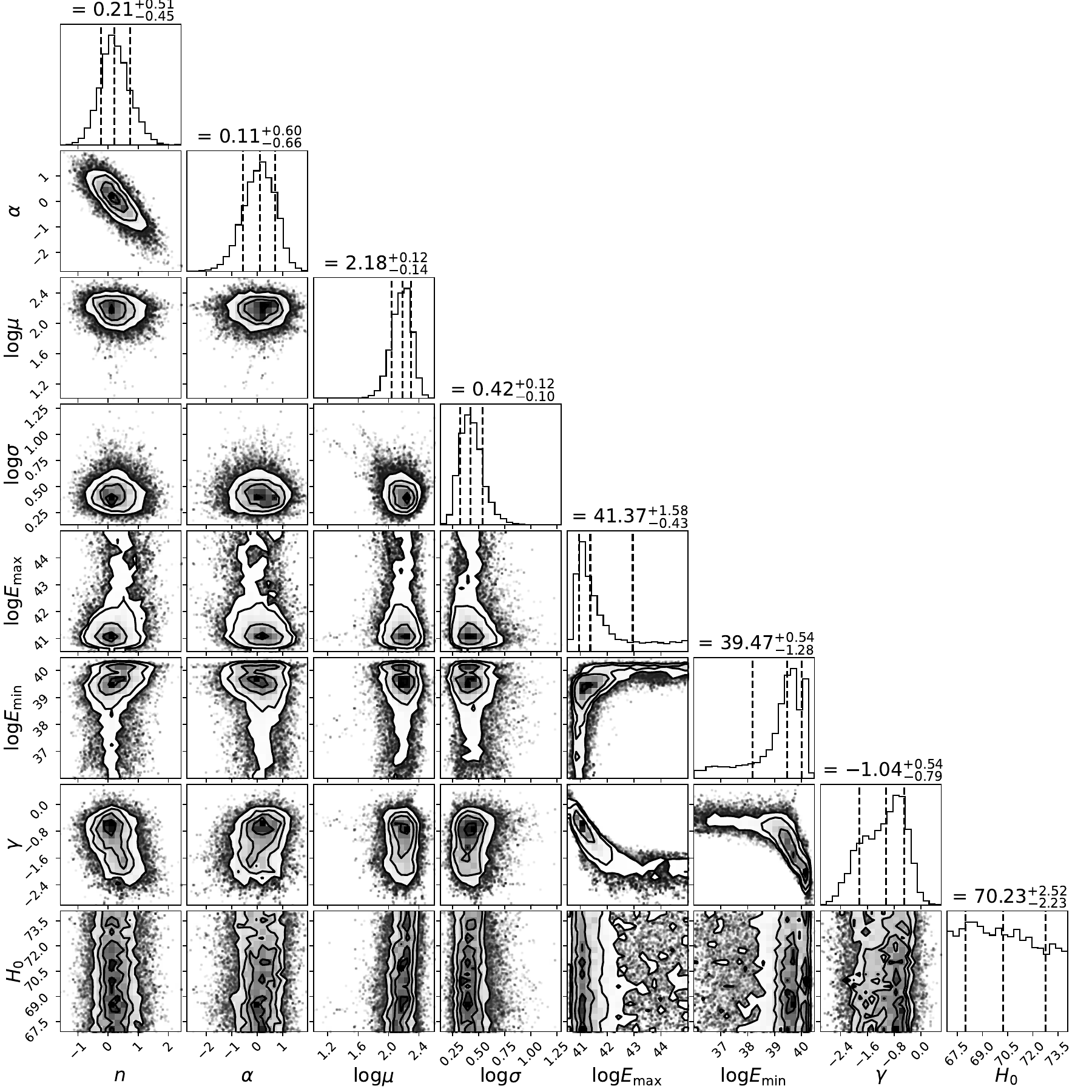}
\caption[]{Results from the MCMC analysis including FAST, DSA and CRAFT FRBs. The parameters are identical to those described in Table~\ref{table:params}.}
\label{fig:MCMCbase}
\end{center}
\vspace{-3ex}
\end{figure*}

\myedit{We focus our analysis on FRB population parameters in this manuscript and so we limit the value of $H_0$ to within 1$\sigma$ of the \citet{SH0ES2021} and \citet{Planck2018} results which are much more precise estimates than our own \citep{james2022b}. We also complete an analysis allowing $H_0$ to vary freely and find a lower value than previously predicted which is discussed in \ref{sec:params}.} The fluctuation parameter, $F$, shows a strong degeneracy with $H_0$ \citep{Baptista2023} and hence we fix it to a value of 0.32 \citep{Macquart2020, Zhang2021}. We consider a case in which $F$ is not fixed in \ref{sec:vary_F}. Figure~\ref{fig:MCMCbase} summarises our numerical results. 

The quoted values give the median and 1\,$\sigma$ deviations (16\% and 84\% quantiles). Most values are consistent with previous results, but we additionally constrain \Emin{}. \myedit{We also note that a flat value of $\alpha = 0.11^{+0.66}_{-0.60}$ is preferred which differs from the previous preference of $\alpha = -0.99^{+0.99}_{-1.01}$ \citep{james2022b}. However, when allowing $H_0$ to freely vary, we recover the previous preference with $\alpha = -0.92^{+0.77}_{-0.94}$ as noted in \ref{sec:systematics}.}

\begin{table}
\begin{center}
\caption{Limits on the uniform priors used in the MCMC analysis. The parameters are as follows: \nsfr{} gives the correlation with the cosmic SFR history; $\alpha$ is the slope of the spectral dependence; $\mu_{\mathrm{host}}$ and $\sigma_{\mathrm{host}}$ are the mean and standard deviation of the assumed log-normal distribution of host galaxy DMs; \Emax{} notes the exponential cutoff of the luminosity function (modelled as a Gamma function); \Emin{} is a hard cutoff for the lowest FRB energy; $\gamma$ is the slope of the luminosity function; and $H_0$ is the Hubble constant. The host parameters $\mu_{\mathrm{host}}$ and $\sigma_{\mathrm{host}}$ are in units of \DMunit{} in log space, \Emax{} and \Emin{} are in units of ergs and $H_0$ is in units of km$\:$s$^{-1}\:$Mpc$^{-1}$. The limits on \Emax{} and \Emin{} were chosen as the distributions are uniform on the extrema of these ranges. \myedit{The limits on $H_0$ were represent a 1\,$\sigma$ interval around the \citet{Planck2018} and \citet{SH0ES2021} results.}}
\label{table:params}
\begin{tabular}{lcc}
\hline
Parameter & Prior Min & Prior Max \\
\hline 
\nsfr{} & -2.0 & 6.0 \\
$\alpha$ & -4.0 & 4.0 \\
$\mu_{\mathrm{host}}$ & 1.0 & 3.0 \\
$\sigma_{\mathrm{host}}$ & 0.1 & 1.5 \\
log$_{10}$(\Emax{}) & 40.5 & 45.0 \\
log$_{10}$(\Emin{}) & 36.0 & 40.5 \\
$\gamma$ & -3.0 & 1.0 \\
$H_0$ & \myedit{66.9} & \myedit{74.08} \\
\hline
\end{tabular}
\end{center}
\end{table} 

\subsection{\Emin{} constraints}
We find \myedit{log\,\Emin{}(erg)$=39.47^{+0.54}_{-1.28}$}. This is the first time such a constraint has been obtained from a $z$--DM analysis. However, studies of strong repeaters have also aimed to probe the minimum FRB energy.

For flat values of $\gamma$ ($> -1$), we obtain no constraint on \Emin{} and the posterior distribution is limited by our prior. This is because the lower energy threshold is not as significant when there is no abundance of low-energy events. As such, the lower bound on our obtained \Emin{} value is not as meaningful as suggested.

\begin{figure}
\begin{center}
\includegraphics[width=\linewidth]{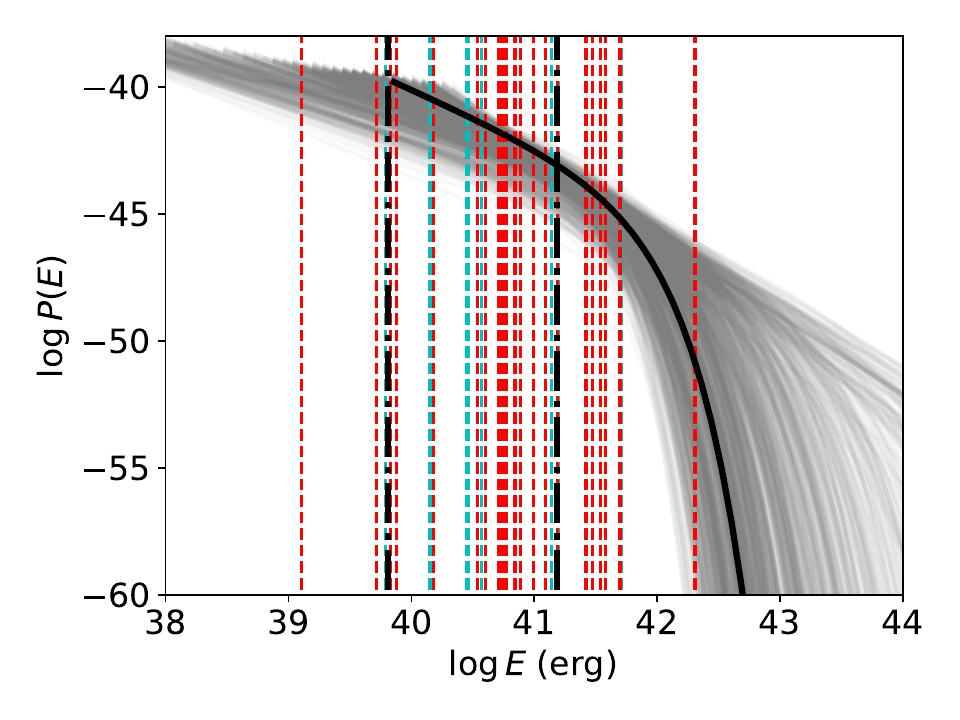}
\caption[]{In grey are 1000 luminosity functions from the MCMC sample. The solid black line shows the best-fit luminosity function. \Emin{} and \Emax{} are shown as black dash-dotted lines. Estimated energies of FRBs with associated redshifts are shown as vertical dashed lines assuming an average beam sensitivity. Those in red were used in the fitting process and those in cyan were not. We do not express it visually, however, each FRB energy also has large uncertainties associated with it due to ambiguities of where the FRB was detected within the beam.}
\label{fig:pE}
\end{center}
\vspace{-3ex}
\end{figure}

Figure~\ref{fig:pE} shows our best-fit luminosity function with the fitted \Emin{} and \Emax{} values given as solid black lines. The estimated energies for FRBs with an associated redshift are also shown (where the cyan lines were not used in the fitting process). The most immediate concern is FRBs having energies below our \Emin{} value which should be a hard cutoff. This is due to the expected energies shown assuming an average value of the beam sensitivity. We do not use any information about where the FRB was detected in the beam and hence determine $P($SNR) via
\begin{equation}
    P(\mathrm{SNR}) = \sum_B P(\mathrm{SNR} \mid B) P(B),
\end{equation}
where $B$ denotes the beam sensitivity. The edges of the beam have orders of magnitude less sensitivity than the centre and hence if these FRBs were detected on the edges, they would have orders of magnitude more intrinsic energy. Thus, FRBs below \Emin{} are allowed at a lower probability by mandating that they are detected on the edge of the beam. If information regarding the exact placement of FRBs in the beam were to be included, such ambiguities could be eliminated and a more stringent \Emin{} could be obtained. We have such information for CRAFT FRBs \citep{Macquart2018a} and hence can include this information in the future, although doing so poses significant computational challenges. For other surveys, making this information public will be of great use in such an analysis.

We also note that our analysis tends to favour the minimum and maximum allowed values when fitting for limiting parameters such as \Emax{} and \Emin{} respectively. The maximum FRB energy was previously predicted to be lower than that of FRB 20220610A \citep{Ryder2023} and was revised with this detection. Likewise, it is likely that as more low-energy events are detected, this limit on \Emin{} will also be constrained to lower values.

\subsection{Comparison of \Emin{} with other literature}
\citet{Agarwal2019} gave constraints on the luminosity function from the non-detection of an FRB when observing the Virgo Cluster. Assuming an all-sky rate of 10$^4$ FRBs per day above a 1\,Jy threshold, they found constraints of $\alpha < 1.52$ and $L_{\mathrm{min}} > 1.6 \times 10^{40}$\,erg\,s$^{-1}$ which approximately corresponds to \Emin{}$> 10^{37}$\,erg\,s given an FRB of width $\sim 1$\,ms. These are consistent with our results.

\citet{Li2021} and \citet{Hewitt2022} used FAST and Arecibo respectively to observe the active repeater FRB 20121102A. In both instances, they observe an increase in bursts below $\sim 10^{38}$\,erg and detect bursts down to $\sim 10^{37}$\,erg which is well below our \Emin{} value. Similarly, \citet{Zhang2022} observed the actively repeating FRB 20201124A with FAST and all of the detected bursts were below our value for \Emin{}.

\myedit{While at face-value our results seem to be in strong contention with these results from strong repeaters, it is more likely that this suggests limitations to our model of the luminosity function. Currently, we use \Emin{} as a hard cutoff as has been previously standardised in the literature. However, the physical interpretation of \Emin{} may not be an actual `minimum energy', but may indicate a lack of low-energy FRBs in comparison to expectation. \citet{Li2021} note a downturn in the rate of bursts between $1 - 3 \times 10^{38}$\,erg which would result in such a lack. While they do see an increase in the burst rate at energies below $\sim 10^{38}$\,erg, this region falls towards the long tail of the \Emin{} posterior distribution which suggests that most surveys are not sensitive to these energies. As such, \Emin{} may be probing this flattening of the spectrum. However, even with such an interpretation, our value is still higher than the break (quoted at $3 \times 10^{38}$\,erg) observed in FRB 20121102A, although it is consistent at the 2\,$\sigma$ level.}

\myedit{Alternatively, strong repeaters may have a different luminosity function to apparently once-off detection events.} Even comparing the aforementioned results of FRBs 20121102A and 20201124A, the luminosity functions of these FRBs differ. Our analysis fits parameters to the entire population and hence is more indicative of the average across all FRBs while such strong repeaters must necessarily be very rare \citep{James2019}. Additionally, while we do include repeating FRBs in our analysis, these FRBs were not detected as repeaters in the surveys that we use. As such, without external information, the surveys view these repeaters as single bursts and so we measure the average values across the population of single burst detections. This discrepancy may therefore suggest that strong repeaters have unique luminosity functions compared to single burst detections.

We also note that a galactic magnetar has produced an FRB-like burst \citep{Bochenek2020} which has been suggested to come from the same sample as extragalactic FRBs. This radio burst had an isotropic equivalent energy of $2.2 \times 10^{35}$\,erg which is significantly below our fitted threshold. Assuming that this burst is from the same sample as extragalactic FRBs also suggests that our results are indicative of a low energy flattening of the spectrum rather than a hard cutoff.

\section{Predictions of $z$ and DM distributions} \label{sec:predictions}
\subsection{FAST sensitivity in $z$--DM space} \label{sec:FASTrates}

\begin{figure}
\centering
\includegraphics[width=\textwidth]{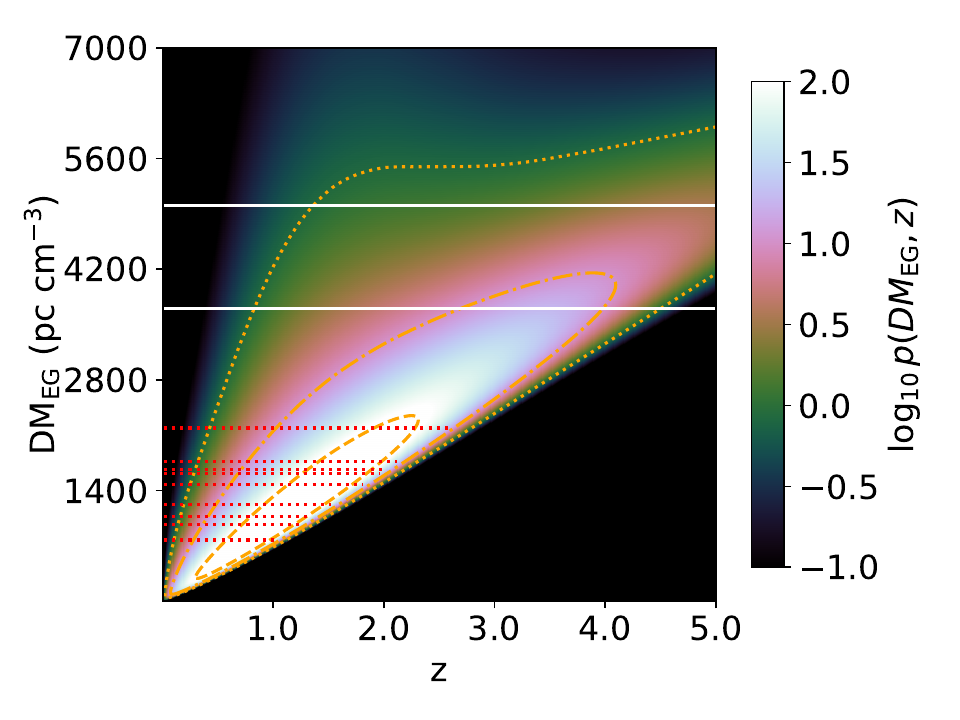}
\caption[]{Predictions of the $z$--\DMEG{} distribution of FAST FRBs using the best-fit parameters for a `default' set of model choices as discussed in \ref{sec:systematics}. The horizontal dashed lines show the expected \DMEG{} values for the 9 unlocalised FRBs after subtracting \DMhalo{} and \DMNE{} from \DMobs{}. The horizontal white lines are the maximum searched DMs of 3700 and 5000\,\DMunit{} for each survey. Shown in orange are the 50\%, 95\% and 99\% probability contours.}
\label{fig:FASTpzdm}
\vspace{-3ex}
\end{figure}

\begin{figure}
\begin{subfigure}{\textwidth}
  \centering
  \includegraphics[width=\textwidth]{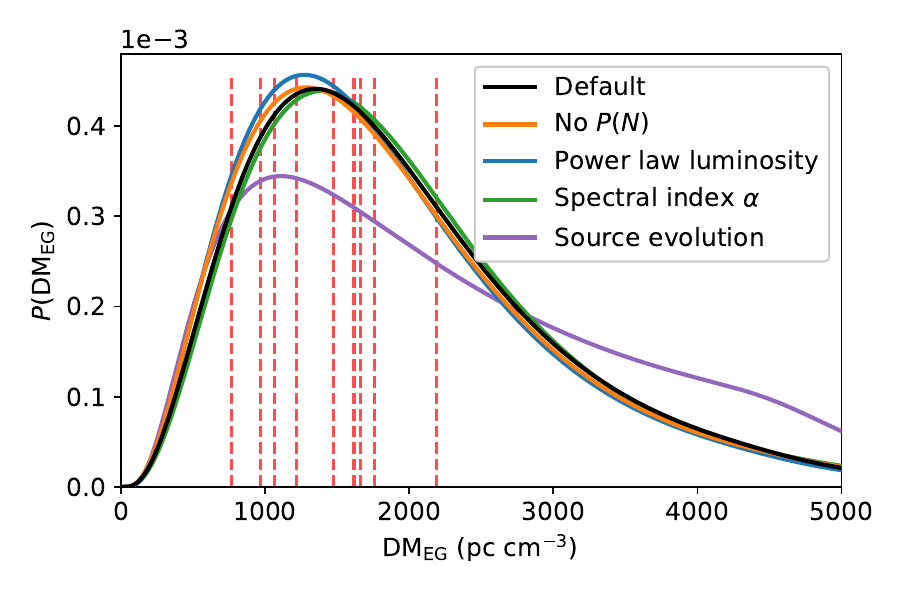}
\end{subfigure}

\begin{subfigure}{\textwidth}
  \centering
  \includegraphics[width=\textwidth]{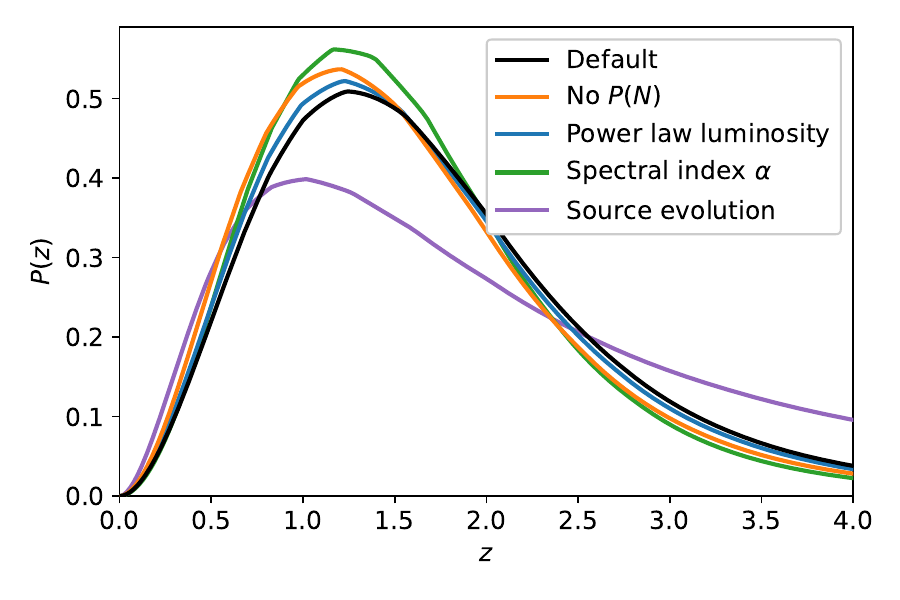}
\end{subfigure}

\caption{The predicted \DMEG{} and $z$ distributions of FAST FRBs. Vertical dashed lines show the estimated \DMEG{} values of the FRBs in this survey which have a typical uncertainty of $50 \sim 200$\,\DMunit\. None of these FRBs have a corresponding $z$. The different colours of solid lines represent different model choices which are mostly arbitrary. These model systematics are discussed in \ref{sec:systematics}.}
\label{fig:FASTpred}
\end{figure}

Figure~\ref{fig:FASTpzdm} shows the predicted $z$--DM distribution of FRBs detected by FAST given the best-fit parameters of our analysis and the default model implementations described in \citet{james2022b}. Figure~\ref{fig:FASTpred} then shows the marginalised distributions for $P$(\DMEG{}) and $P(z)$ with these default model choices and using alternative model choices as described in \ref{sec:systematics}. The dashed vertical lines show the expected \DMEG{} values of the 9 FRBs from FAST used in this analysis. That is, we place the vertical lines at \DMEG{} = \DMobs{} - \DMNE{} - \DMhalo{}. The range of percentages we quote hereafter corresponds to uncertainties due to the model choices discussed in \ref{sec:systematics}. It does not account for the uncertainties that we have in each parameter.

The 9 FRBs that FAST has detected within the two published surveys had a surprisingly large average DM in comparison to other surveys. However, we find that these DMs are consistent with our analysis, serving as a method of ratification. For the 4 FRBs detected in the CRAFTS survey, a maximum search DM of 5000\,\DMunit{} was used \citep{Niu2021} which excludes 1 -- 4\% of possible detections. The GPPS survey used a maximum search DM of 3700\,\DMunit{} which excludes 7 -- 18\% of possible detections.

Currently, none of these FRBs have been localised to a host galaxy and hence nothing is known about the empirical $z$ distribution of FAST FRBs. We note that FAST has detected FRB 20190520B which was later localised with the Jansky Very Large Array to a host galaxy at $z=0.241$ \citep{Niu2022}, however, we do not know the survey parameters of this detection and hence we cannot include it in this analysis. Here, we provide our prediction for the FAST $z$ distribution. The distribution peaks at $z \approx 1$ and has 73 -- 77\% of FRBs detected beyond $z\gtrsim1$. Currently, FRB 20220610A is the furthest FRB that has been localised to a host galaxy system, at $z \approx 1$ \citep{Ryder2023, Gordon2024} and thus FAST will probe an entirely new region of the parameter space. 

We also predict that 25 -- 41\% of FAST FRBs will be detected beyond $z\gtrsim2$. At $z \approx 2$, the empirical SFR model of \citet{Madau2014} that we use turns over. Thus, this high-$z$ region may allow us to differentiate between FRB source evolution following SFR or not (see \ref{sec:SFR} for further discussion).

FRBs probe the content of ionised gas in the Universe and hence may be able to probe epochs such as the reionisation of He II  . This reionisation is expected to occur at $z \approx 3$ \citep{He2Worseck2016, He2Worseck2019} and hence FRBs that come from beyond this could detect this epoch via a break in the Macquart relation \citep{Macquart2020}. No other telescope that we analyse is expected to detect FRBs close to $z=3$, however, our predictions show that 6 -- 20\% of detections will be beyond $z\gtrsim3$ for FAST.

While detecting FRBs out to these large redshifts is certainly exciting, the localisation ability of FAST is not sufficient to robustly associate the FRB to a single host galaxy for any of these redshifts and hence $z$ values cannot be obtained unless the FRB is seen to repeat and can be localised by other instruments. Additionally, FRBs at such large redshifts have a significantly higher probability of intersecting intervening halos and hence it will become increasingly valuable to include data from observational schemes such as the FLIMFLAM survey \citep{Khrykin2024} in future analyses.

\begin{figure}
\begin{center}
\includegraphics[width=\textwidth]{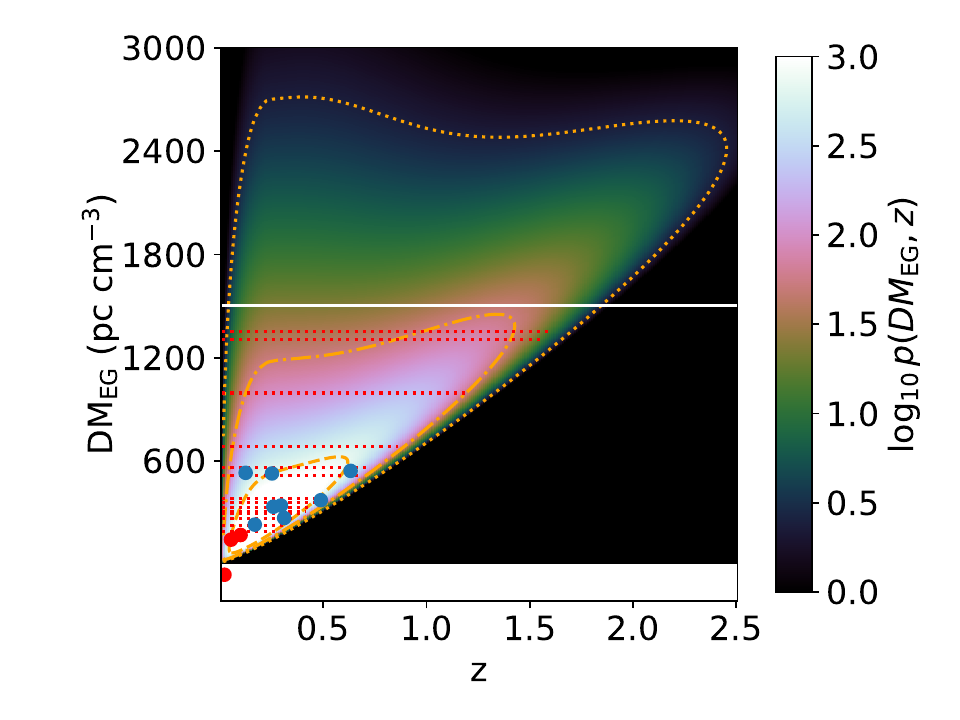}
\caption[]{Predictions of the $z$--\DMEG{} distribution of DSA FRBs using the best-fit parameters for a `default' set of model choices as discussed in \ref{sec:systematics}. The horizontal dashed lines show the expected \DMEG{} values for the unlocalised FRBs after subtracting \DMhalo{} and \DMNE{} from \DMobs{}. The points show localised FRBs. Red points are used in the fitting process while the blue points only utilise DM information (see Section \ref{sec:DSA}). The horizontal white line is the maximum searched DM of 1500\,\DMunit{}. Shown in orange are the 50\%, 95\% and 99\% probability contours. The white strip at the bottom corresponds to a negative \DMEG{} as the assumed \DMEG{} of FRB 20220319D is negative.}
\label{fig:DSApzdm}
\end{center}
\vspace{-3ex}
\end{figure}

\begin{figure}
\begin{subfigure}{\textwidth}
  \centering
  \includegraphics[width=\textwidth]{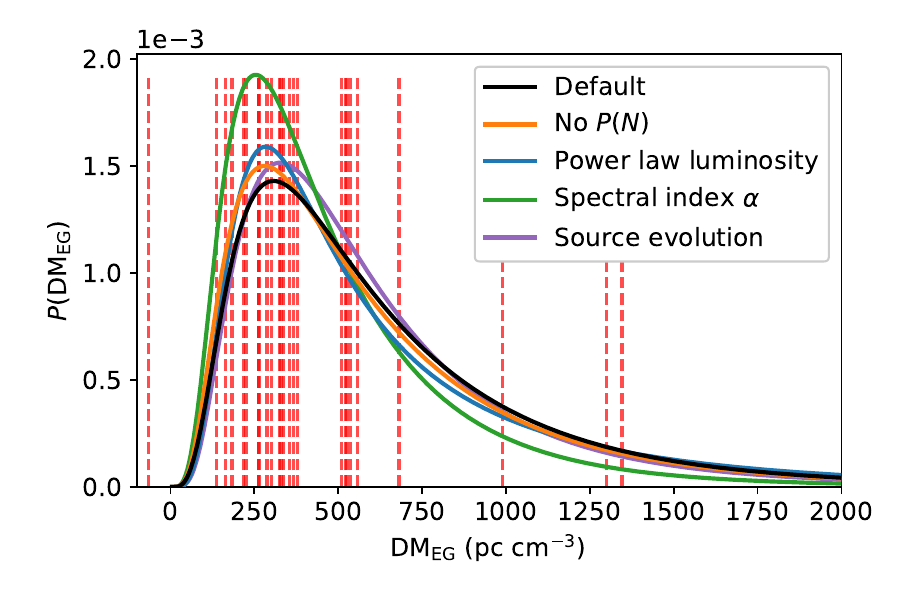}
\end{subfigure}

\begin{subfigure}{\textwidth}
  \centering
  \includegraphics[width=\textwidth]{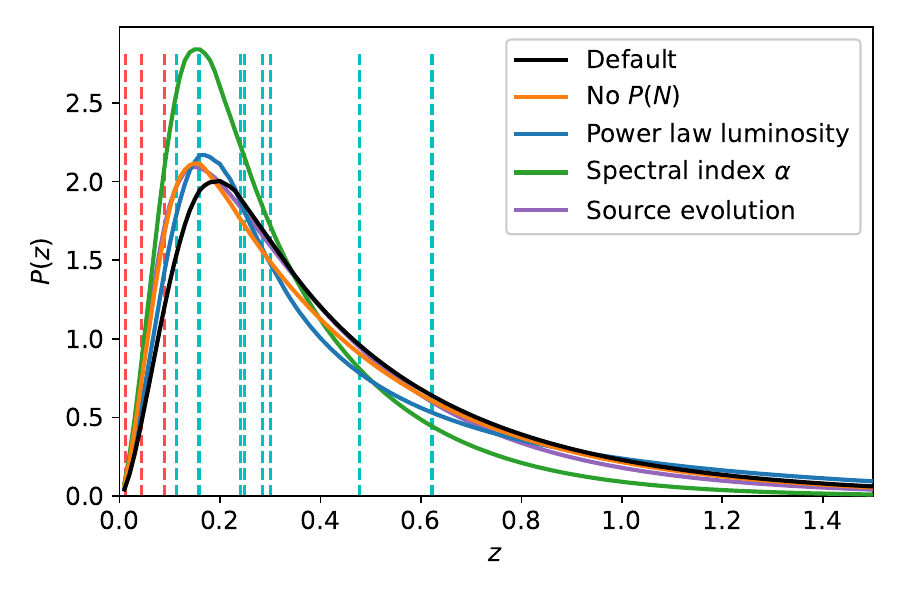}
\end{subfigure}

\caption{The predicted \DMEG{} and $z$ distributions of DSA FRBs. Vertical dashed lines show the estimated \DMEG{} values which have a typical uncertainty of $\sim 100$\,\DMunit\ and the observed $z$ values of the FRBs in this survey. For the $P(z)$ distribution, the red dashed lines show localisations that are used in the fitting process while the $z$ values of the blue dashed lines have not been used. The different colours of solid lines represent different model choices which are mostly arbitrary. These model systematics are discussed in \ref{sec:systematics}.}
\label{fig:DSApred}
\end{figure}

\begin{figure}
\begin{center}
\includegraphics[width=\textwidth]{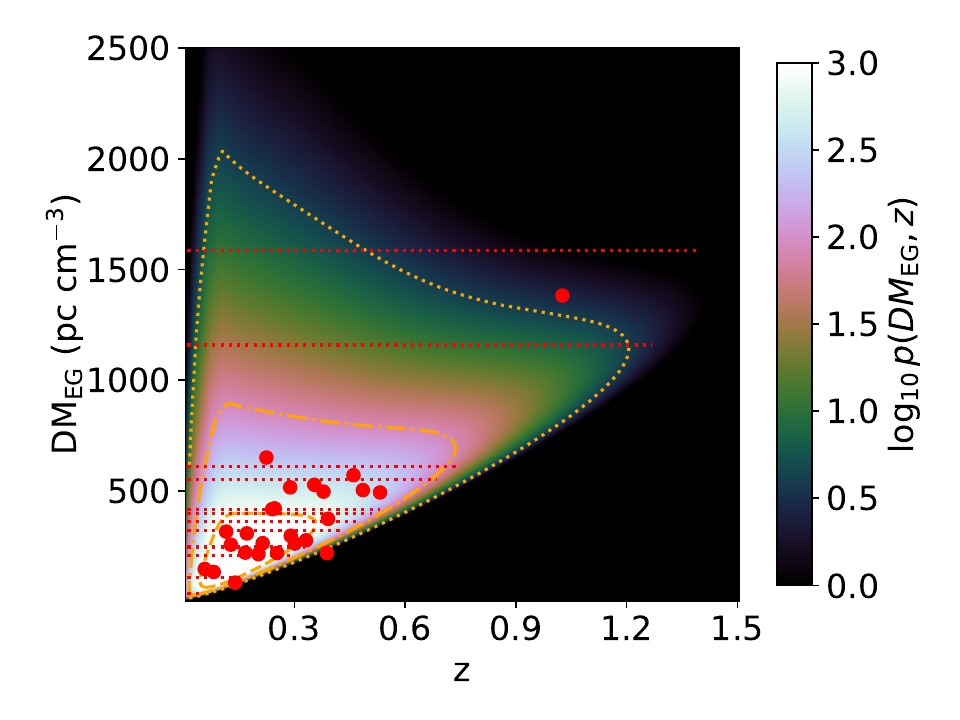}
\caption[]{Predictions of the $z$--\DMEG{} distribution of CRAFT/ICS FRBs averaged over the three frequency groups and using the best-fit parameters for a `default' set of model choices as discussed in \ref{sec:systematics}. The horizontal dashed lines show the expected \DMEG{} values for the unlocalised FRBs after subtracting \DMhalo{} and \DMNE{} from \DMobs{}. The points show localised FRBs. Shown in orange are the 50\%, 95\% and 99\% probability contours.}
\label{fig:CRAFTpzdm}
\end{center}
\vspace{-3ex}
\end{figure}

\subsection{DSA and CRAFT/ICS sensitivity in $z$--DM space}

Similarly to our results for FAST, Figure~\ref{fig:DSApzdm} shows the $z$--DM distribution for DSA FRBs using our best-fit parameters of the default model and Figure~\ref{fig:DSApred} shows the marginalised distributions. We do not use the redshifts of the vertical blue dashed lines in the fitting process as justified in Section \ref{sec:DSA}. For a comparison, we show the $z$--DM distribution averaged over the three CRAFT/ICS surveys in Figure~\ref{fig:CRAFTpzdm}. 

In regards to the redshift distribution, we predict that 2 -- 12\% of DSA FRBs will be detected with $z \gtrsim 1$ and 0.02 -- 1.7\% with $z \gtrsim 2$. For CRAFT/ICS, the fraction of FRBs we predict with $z \gtrsim 1$ is 0.05 -- 8\%. We expect DSA to be more comparable to the upgraded CRAFT detection system (CRACO; \bannisterinprep, \wanginprep) which is in the commissioning stages. Therefore, we expect both DSA and CRAFT/CRACO to produce many localisations in the local Universe and out beyond the current limits of the CRAFT/ICS detection system.

DSA has been using a maximum searched DM of 1500\,\DMunit{} during the commissioning observations \citep{Law2023}. With such a limit, we predict that 3 -- 8\% of FRBs are missed, which will preferentially be high-$z$ FRBs. The CRAFT/ICS detection system \citep[FREDDA;][]{Bannister2019} searches up to 4096 time samples. This corresponds to maximum search DMs of
\begin{itemize}
    \item 1046\,\DMunit{} for \icslow{},
    \item 3468\,\DMunit{} for \icsmid{} and
    \item 7428\,\DMunit{} for \icshigh{}.
\end{itemize}
Thus, \icslow{} misses 1 -- 2\% of possible detections, \icsmid{} misses 0.03 -- 0.07\% and \icshigh{} misses a negligible amount.

In general, these two instruments will provide a complementary sample of localised FRBs to the existing localisations and will hopefully begin to fill the more distant redshift range around $z \sim 1$. It is these numerous localisations that will allow for analyses such as this one to produce more refined results and will allow FRBs to shed light on cosmological issues such as the Hubble tension.

\section{Conclusion} \label{sec:conclusions}
We present our latest results fitting FRB population parameters and $H_0$ following on from the work of \citet{james2022b}. We include additional CRAFT/ICS, FAST and DSA FRBs and make two primary improvements to the analysis. Firstly, we implement an MCMC sampler to ease computational strain which allows the additional parameter of \Emin{} to be fit. Secondly, we implement an uncertainty in \DMG{}, separately for its ISM and halo components.

We obtain the first constraint on \Emin{} from a $z$--DM analysis of \myedit{log\,\Emin{}(erg)=39.47$^{+0.54}_{-1.28}$}. This is much higher than expected and exceeds energies detected in strong repeaters by orders of magnitude. We expect that this is suggestive of a break in the luminosity function at low energies, although it may suggest that apparently single bursts have a unique luminosity function in comparison to repeaters. We also note that FRBs in the survey have an expected energy less than this value, however, they are still allowed by accounting for the possibility that they are detected on the edges of the beams. We aim to include the exact beam sensitivity of detection in future versions of the code.

With the inclusion of FAST, the parameter space that is open to FRB science increases considerably. We are able to probe significantly larger redshifts in which new science becomes available. We predict that a majority of FAST FRBs will be detected beyond the current highest redshift FRB at $z \sim 1$ and 25 -- 41\% will be detected beyond $z \gtrsim 2$. While these prospects could allow us to probe FRB progenitor evolution and He II ionisation, it is currently limited by the ability to localise these FRBs sufficiently to obtain an associated redshift. With the current maximum search DM of 3700\,\DMunit{}, we predict that FAST is missing 7 -- 18\% of possible FRB detections. With the previous limit of 5000\,\DMunit{}, we predict that they only miss 1 -- 4\%.

DSA probes a similar parameter space to CRAFT surveys but is more sensitive and is expected to obtain a large number of FRBs with corresponding redshifts. These FRBs will help reduce the statistical noise of our analysis and are what is needed to shed light on the Hubble tension. We predict that of order 2 -- 12\% of their FRBs will be detected with $z \gtrsim 1$ in comparison to the 0.05 -- 8\% from CRAFT/ICS surveys. We also predict that DSA will miss 3 -- 8\% of possible FRBs given their maximum search DM of 1500\,\DMunit{}.

\begin{acknowledgement}
We thank Nicolas Tejos for his comments on the manuscript \myedit{and the referee for their help in clarifying the message of the paper}. This work was performed on the OzSTAR national facility at Swinburne University of Technology. The OzSTAR programme receives funding in part from the Astronomy National Collaborative Research Infrastructure Strategy (NCRIS) allocation provided by the Australian Government. This scientific work uses data obtained from Inyarrimanha Ilgari Bundara, the CSIRO Murchison Radio-astronomy Observatory. We acknowledge the Wajarri Yamaji as the Traditional Owners and native title holders of the Observatory site. CSIRO’s ASKAP radio telescope is part of the Australia Telescope National Facility. The operation of ASKAP is funded by the Australian Government with support from the National Collaborative Research Infrastructure Strategy. ASKAP uses the resources of the Pawsey Supercomputing Research Centre. The establishment of ASKAP, Inyarrimanha Ilgari Bundara, the CSIRO Murchison Radio-astronomy Observatory and the Pawsey Supercomputing Research Centre are initiatives of the Australian Government, with support from the Government of Western Australia and the Science and Industry Endowment Fund.

\end{acknowledgement}

\paragraph{Funding Statement}
This research was supported by an Australian Government Research Training Program (RTP) Scholarship.
CWJ and MG acknowledge support through Australian Research Council (ARC) Discovery Project (DP) DP210102103. 
RMS acknowledges support through Australian Research Council Future Fellowship FT190100155 and Discovery Project DP220102305. 
A.C.G. and the Fong Group at Northwestern acknowledges support by the National Science Foundation under grant Nos. AST-1909358, AST-2308182 and CAREER grant No. AST-2047919. 
J.X.P., A.C.G. acknowledge support from NSF grants AST-1911140, AST-1910471 and AST-2206490 as members of the Fast and Fortunate for FRB Follow-up team. 
ATD acknowledges support through Australian Research Council Discovery Project DP22010230.

\paragraph{Competing Interests}
None.

\paragraph{Data Availability Statement}
The code and data used to produce our results can be found at \href{https://github.com/FRBs/zdm}{https://github.com/FRBs/zdm}.

\printendnotes

\printbibliography 

\appendix
\section{Additional parameter discussion} \label{sec:other_params}
\myedit{In the main body we restrict the value of $H_0$ to values consistent with existing literature. Here we extend the work presented in the main body and allow $H_0$ to vary freely for a more complete view of the parameters of the model and their interdependencies. We present our expected parameter constraints and discuss parameter correlations. These results are presented in Table \ref{table:systematics}.}

\subsection{Population parameter constraints} \label{sec:params}
Most parameter values are consistent with what was obtained in previous analyses. The most substantial difference was when $H_0$ was allowed to vary freely and in this instance, $H_0$ decreased from $73^{+12}_{-8}$\,km\,s$^{-1}$\,Mpc$^{-1}$ to $58^{+13}_{-9}$\,km\,s$^{-1}$\,Mpc$^{-1}$, although this change is in the context of a fixed $F$ value. We find that the lower limit is shifted downwards, however, the uncertainties are large enough that our results are still consistent with previous values. The lower limit of $H_0$ is mostly determined by the sharp drop-off in $P(z,$DM) for objects with low DM for their corresponding redshift. By allowing for uncertainty in \DMISM{} and \DMhalo{}, we ascribe an uncertainty to \DMEG{} and hence allow for the possibility that FRBs are further from this `cliff', thereby weakening the strong constraint that they give. \myedit{Thus, accurately knowing \DMISM{} is important in constraining a value of $H_0$.}

In addition to $H_0$, we see that both $\mu_{\mathrm{host}}$ and $\sigma_{\mathrm{host}}$ decreased. This is partially due to their correlation with $H_0$ but also due to uncertainties in \DMG{} previously being absorbed into these terms which we now model explicitly. Our results for the mean host contribution of $105^{+61}_{-51}$\,\DMunit{} is in good agreement with the value obtained from the FLIMFLAM survey of $90^{+29}_{-19}$\,\DMunit{} \citep{Khrykin2024}.

In general, we find that our uncertainties are comparable to or up to 50\% larger than what was previously obtained. The introduction of uncertainty on \DMG{} naturally increases the uncertainty on parameters, while the introduction of more data from FAST, DSA and recent CRAFT/ICS surveys results in decreased statistical uncertainties.

\subsection{Parameter correlations}
Parameter correlations can be seen in Figure~\ref{fig:MCMCbase}. The \nsfr{} parameter describes the relation between the SFR history of the Universe and the number of FRB progenitors and hence indicates the cosmic source evolution. Current FRB detections have probed up to $z \sim $1 \citep{Ryder2023}, in which the \citet{Madau2014} SFR model that we use shows SFR consistently increasing with $z$. A higher \nsfr{} value therefore suggests an increased number of FRBs in the distant Universe. Thus, correlations with \nsfr{} indicate how a particular parameter modifies the ratio between the number of nearby and distant FRBs that one expects to detect.

As discussed in \ref{sec:rates}, a steeper $\alpha$ biases detections towards the nearby Universe given a consistent observing band. This produces a strong anti-correlation between \nsfr{} and $\alpha$. As such, we note that for any given $\alpha$ value, \nsfr{} is tightly constrained. However, these two parameters are highly degenerate. A more detailed discussion of this degeneracy is presented in \citet{james2022}.


The three luminosity function parameters \Emax{}, \Emin{} and $\gamma$ also show strong correlations. \Emax{} and $\gamma$ show a negative correlation which intuitively is derived from limiting the number of high-energy events. A flatter $\gamma$ allows for a greater relative number of high-energy FRBs therefore requiring a lower \Emax{} to remain consistent with the observed population. 

Similarly, \Emin{} and $\gamma$ also exhibit a negative correlation. The value of \Emin{} is more impactful at low $z$ as this is the region in which detections are limited by the intrinsic burst energy. In the high-$z$ regime, the observed fluence of a burst is naturally decreased and hence the SNR threshold is more limiting. Thus, a higher value of \Emin{} limits the number of low-$z$ FRBs which requires a steeper (more negative) $\gamma$ to stay consistent with the observed population. \Emin{} and \Emax{} are therefore positively correlated via their correlations with $\gamma$.

We also note that a steeper $\gamma$ suppresses high-energy events and therefore does not require a high-energy cutoff. On the converse, a flatter $\gamma$ suppresses low-energy events but allows for many more high-energy events. As such, steep $\gamma$ values tightly constrain \Emin{} but give rise to a long tail in \Emax{} while flat $\gamma$ values tightly constrain \Emax{} and allow for a long tail in \Emin{}. Due to this, we place an artificial lower limit on \Emin{} and an artificial upper limit on \Emax{} with the assigned priors.

Correlations with $H_0$ are discussed in \citet{james2022b} and hence we do not repeat them here. The only new parameter introduced is \Emin{} which exhibits a weak positive correlation with $H_0$. This is because a higher \Emin{} biases detections to higher redshifts for a given DM while a higher $H_0$ does the inverse.

\section{Alternative model choices} \label{sec:systematics}
There are a number of assumptions that have been made in our analysis. As the emission mechanism of FRBs is not known, it is difficult to find physical justifications for many models. As such, there are ambiguities in the choice of model implemented. In this section, we discuss model systematics stemming from the exclusion of $P(N)$, the interpretation of $\alpha$, the choice of luminosity function, the assumption of source evolution following the cosmic SFR and uncertainty in the value of the fluctuation parameter, $F$. A comparison of results when including each of these model variations is presented in Table~\ref{table:systematics}. 

These systematics cause the Hubble constant to vary within $\pm 2$\,km\,s$^{-1}$\,Mpc$^{-1}$ (with the exclusion of varying $F$ as this parameter is degenerate with $H_0$) which is substantially less than the statistical uncertainties. However, these systematics will be relevant in the eventuation of resolving the Hubble tension which would require an accuracy close to 1\,km\,s$^{-1}$\,Mpc$^{-1}$ \citep{SNOWMASS2022}. Thus, while we cannot distinguish between the models at present, this will be important in future studies. 

\begin{table*}
\begin{center}
\caption{Parameter constraints from the MCMC analysis when including FAST, DSA and CRAFT FRBs. The constraints we quote give the median value and the corresponding uncertainties are the 16\% and 84\% quantiles taken from analogous plots to Figure~\ref{fig:MCMCbase}. Given is the parameter name and constraints (1) with the default analysis parameter, (2) when ignoring $P(N)$, (3) when using a spectral index interpretation of $\alpha$, (4) when using a power law luminosity function, (5) when assuming the source evolution does not follow SFR and (6) allowing the fluctuation parameter, $F$, to vary. We also give the previous results from \citet{james2022b}. The host parameters $\mu_{\mathrm{host}}$ and $\sigma_{\mathrm{host}}$ are in units of \DMunit{} in log space, \Emin{} and \Emax{} are in units of erg and $H_0$ is in units of km$\:$s$^{-1}\:$Mpc$^{-1}$.}
\label{table:systematics}
\begin{tabular}{lccccccc}
\hline
Parameter & Default & No $P(N)$ & Spectral index $\alpha$ & Power-law & No SFR correlation & Varying $F$ & James et al. (2022) \\
\hline 
\nsfr{} & 0.91$^{+0.61}_{-0.55}$ & $^a$1.75$^{+0.79}_{-0.83}$ & 0.72$^{+0.45}_{-0.46}$ & 0.93$^{+0.53}_{-0.56}$ & 0.17$^{+0.41}_{-0.40}$ & 1.32$^{+0.80}_{-0.91}$ & 1.13$^{+0.49}_{-0.41}$ \\
$\alpha$ & -0.92$^{+0.77}_{-0.94}$ & $^b$-- & -0.69$^{+0.54}_{-0.54}$ & -1.10$^{+0.75}_{-0.87}$ & -0.30$^{+0.77}_{-0.92}$ & -2.31$^{+1.62}_{-1.69}$ & -0.99$^{+0.99}_{-1.01}$ \\
$\mu_{\mathrm{host}}$ & 2.02$^{+0.20}_{-0.29}$ & 1.98$^{+0.22}_{-0.35}$ & 1.98$^{+0.24}_{-0.40}$ & 2.04$^{+0.19}_{-0.30}$ & 2.08$^{+0.21}_{-0.30}$ & 2.06$^{+0.17}_{-0.22}$ & 2.27$^{+0.12}_{-0.13}$ \\
$\sigma_{\mathrm{host}}$ & 0.46$^{+0.17}_{-0.13}$ & 0.44$^{+0.20}_{-0.13}$ & 0.47$^{+0.27}_{-0.14}$ & 0.42$^{+0.17}_{-0.11}$ & 0.44$^{+0.19}_{-0.11}$ & 0.49$^{+0.16}_{-0.13}$ & 0.55$^{+0.12}_{-0.09}$ \\
$^c$log$_{10}$(\Emax{}) & 41.42$^{+0.94}_{-0.41}$ & 41.35$^{+0.88}_{-0.34}$ & 41.67$^{+1.01}_{-0.50}$ & $^d$41.91$^{+2.23}_{\mathrm{--}}$ & 41.29$^{+0.67}_{-0.32}$ & 41.51$^{+1.42}_{-0.48}$ & 41.26$^{+0.27}_{-0.22}$ \\
$^c$log$_{10}$(\Emin{}) & 39.49$^{+0.39}_{-1.48}$ & 39.30$^{+0.39}_{-1.68}$ & 39.58$^{+0.45}_{-1.85}$ & 39.74$^{+0.17}_{-0.23}$ & 39.12$^{+0.55}_{-1.78}$ & 39.64$^{+0.42}_{-1.56}$ & -- \\
$\gamma$ & -1.16$^{+0.57}_{-0.68}$ & -1.06$^{+0.49}_{-0.59}$ & -1.17$^{+0.52}_{-0.56}$ & -1.70$^{+0.22}_{-0.19}$ & -0.87$^{+0.37}_{-0.56}$ & -1.30$^{+0.71}_{-0.65}$ & -0.95$^{+0.18}_{-0.15}$ \\
$H_0$ & 58$^{+13}_{-9}$ & 55$^{+13}_{-8}$ & 55$^{+15}_{-8}$ & 58$^{+13}_{-9}$ & 58$^{+19}_{-10}$ & 64$^{+15}_{-13}$ & 73$^{+12}_{-8}$ \\
\hline
\end{tabular}

\floatfoot{$^a$ The \nsfr{} parameter is strongly correlated with $\alpha$ and thus $\alpha$ having no constraint implies that we cannot constrain \nsfr{} either. The quoted \nsfr{} value when not including $P(N)$ is mostly based on the priors of $\alpha$ that we use.
\\$^b$ When not including $P(N)$ we obtain no upper limit and hence no meaningful constraint for $\alpha$. 
\\$^c$ \Emin{} and \Emax{} have long lower and upper tails respectively which do not converge. As such, the quantiles quoted may not accurately reflect the distribution. 
\\$^d$ No lower limit is given for \Emax{} when using a power law luminosity function as this distribution has a sharp cutoff below the mode.}
\end{center}
\end{table*} 

\subsection{Is $P(N)$ reliable?} \label{sec:rates}
The analysis of \citet{james2022b} determines the likelihood of a given cosmological state by calculating $P(z,$ \DMobs{}, SNR) for each FRB and $P(N)$ for each survey, where $N$ is the number of observed FRBs. The expected number of FRBs is calculated by taking an integral of the FRB event rate in $z$--DM space for a given survey and multiplying it by the effective observational time $T_{\mathrm{obs}}$. We normalise the rates by taking a maximum likelihood over all surveys to find the best fit between the expected and observed number of FRBs. 

Table~\ref{table:rates} shows the expected number of FRBs and the observed number of FRBs for each survey given best-fit parameters. We find that the CRAFT/ICS and FAST samples detect a factor of 2 fewer FRBs than expected from \flyseye{} predictions. \citet{ICS2024} note that the ASKAP/ICS sample has a lack of low SNR FRBs which suggests that there is a selection bias against low SNR events. However, even with an increased SNR threshold of 14, the CRAFT/ICS surveys have low rates relative to the rate observed for \flyseye{} observations. 

One possible explanation for this is the presence of radio-frequency interference (RFI) decreasing the effective amount of observing time. RFI is spasmodic and unpredictable, thus making it difficult to characterise $T_{\mathrm{obs}}$. RFI can prevent detections of FRBs but will not artificially create FRBs and hence we expect the \flyseye{} rates to be more accurate. Ultimately, the source of such discrepancies is unknown and hence it is unclear as to whether $P(N)$ is contributing meaningfully to the analysis or is instead introducing inaccuracies. 

\begin{table}
\begin{center}
\caption{Expected and observed number of FRBs in each survey considered when including and excluding $P(N)$. These numbers are only for periods where we have a good estimate of $T_\mathrm{obs}$, while in general, we include more FRBs in the analysis. The total observation time of DSA is unknown and hence this cannot be included in the analysis.}
\label{table:rates}
\begin{tabular}{lcccc}
\hline
Survey & $T_\mathrm{obs}$ (days) & \multicolumn{2}{c}{Expected} & Observed \\
 & & $P(N)$ & No $P(N)$ & \\
\hline
DSA & -- & -- & -- & 25 \\
\FAST & 108.4 & 13.7 & 17.6 & 9 \\
\flyseye & 1274.6 & 11.0 & 10.2 & 20 \\
\icslow & 317.3 & 15.1 & 14.2 & 11 \\
\icsmid & 165.5 & 9.1 & 8.4 & 5 \\
\icshigh & 50.9 & 1.4 & 1.3 & 1 \\
\parkes & 164.4 & 9.3 & 8.6 & 12 \\
\hline
\end{tabular}
\end{center}
\end{table}

Most parameters do not show significant differences with the exclusion of $P(N)$. The exceptions to this are $\alpha$ and \nsfr{}. The upper limit of $\alpha$ shows no constraint and as discussed previously, \nsfr{} is highly correlated with $\alpha$. Thus, the large change in \nsfr{} is likely due to the lack of constraint on $\alpha$ and the priors we select.

Due to cosmic expansion, events with a higher $z$ have a higher intrinsic emission frequency given they are observed in the same band. A steeper $\alpha$ value thus penalises high $z$ events which decreases the expected number of FRBs for more sensitive surveys. As such, a flat $\alpha$ results in an expectation for FAST to detect significantly more FRBs. $P(N)$ thus places an upper limit on $\alpha$ and it is therefore important to have robust estimates for the event rates of each survey.

\subsection{Interpretation of $\alpha$}

Some FRBs are detected with partial band occupancy \citep[e.g.][]{Law2017} and hence the traditional interpretation of spectral index is not applicable. As such, spectral dependence in the FRB population indicates that either there are more low-frequency FRBs (rate interpretation) or low-frequency FRBs on average contain more energy (spectral index interpretation). We use a rate interpretation by default for consistency with \citet{james2022b}.

\citet{Macquart2019} studied ASKAP-detected FRBs to determine the average present in each frequency bin. From this study, they obtained a spectral index of $\alpha=-1.5^{+0.2}_{-0.3}$ which is higher than the value of $\alpha$ that we obtain in any case. This value is still in agreement with the rate interpretation results within $1 \sigma$ and with the spectral index interpretation at $2 \sigma$. When using a spectral index interpretation, $\alpha$ and \nsfr{} both decrease.

\subsection{Power law luminosity function}
As the emission mechanism of FRBs is not known, it is not possible to derive a luminosity function corresponding to the physical processes generating the FRB. Nevertheless, most physical processes can be approximated as a power law to first order and thus \citet{james2022} originally implemented a power law luminosity function with a hard cut-off at the lower and upper energy limits of \Emin{} and \Emax{}. \citet{james2022b} later implemented an exponential cutoff for the upper limit in the form of a Gamma function while still retaining a hard cutoff at \Emin{}. This is the model we have used thus far in our analysis, however, we also recalculate results using a simple power law luminosity function in our model systematics.

The meaning of $E_{\rm max}$ differs between the Gamma function (where it is the downturn energy) and a pure power-law (where it is a sharp cutoff). We thus expect fitting a pure power-law to find a greater value of $E_{\rm max}$ and steeper value of $\gamma$. This is indeed the case (see Table~\ref{table:systematics}). However, the effect on other parameters is slight. 

\subsection{Source evolution dependence on SFR} \label{sec:SFR}

It is still unclear as to whether the evolution of FRB progenitors, $\Phi(z)$ (bursts per proper time per comoving volume), correlates with the SFR history of the Universe ($\Phi(z)~\propto~$SFR$(z)^n$) that peaks close to $z \sim 2$ \citep{Madau2014} or not. Currently, we assume a correlation with SFR, however, \citet{Lin2024} recently noted that they find more consistency with time delay models than an SFR history model using the CHIME gold sample. While testing all models is not feasible, we consider a model in which the event rates scales with a simple power of $\Phi(z) \propto (1+z)^{2.7n}$. However, this model only differs significantly from the SFR history model at higher redshifts beyond $z \sim 1$, which is the furthest confirmed FRB redshift to date \citep{Ryder2023}. As such, we cannot definitively differentiate between the two models until higher redshift objects are found. In general, FAST probes to much higher redshifts than all other surveys and hence is the clearest way to differentiate between these models (see Section \ref{sec:FASTrates} for further discussion).

When assuming the event rate scales with a simple power of $(1+z)^{2.7 n}$, we see a distinct decrease in \nsfr{}, $\alpha$ and $\gamma$. Mathematically, we expect a decrease in \nsfr{} (corresponding to a flattening of the evolution model) as we no longer consider a turnover at $z \sim 2$. This explains the observed decrease from \nsfr{}~$= 0.91^{+0.61}_{-0.55}$ to \nsfr{}~$= 0.17 ^{+0.41}_{-0.40}$. While we expect FAST to have the greatest impact on this, removing FAST from the analysis yields similar results and hence we expect that the lack of localisations prevent FAST from being constraining. The decrease in $\alpha$ and $\gamma$ can then be explained through their correlations with \nsfr{}.

\subsection{Varying F} \label{sec:vary_F}
The fluctuation parameter, $F$, details the amount of scatter in the Macquart relation \citep{Macquart2020}. It varies from 0 to 1 where $F=0$ corresponds to no scatter and $F=1$ is the maximum scatter. Recently, \citet{Baptista2023} showed that $F$ is largely degenerate with $H_0$ and hence we must precisely know $F$ to obtain a strong constraint on $H_0$ and vice versa. In the analysis of this manuscript thus far, we use a constant value of $F=0.32$ \citep{Macquart2020, Zhang2021}. Here, we discuss implications when allowing $F$ to vary freely with a log uniform prior of log\,$F \in U(-2,0)$. 

We obtain a best-fit value of log\,$F = -0.27^{+0.38}_{-0.19}$ which is significantly less constraining than the result of \citet{Baptista2023} which placed a 3\,$\sigma$ lower bound of log\,$F > -0.86$. Nevertheless, we still obtain a constraint on the lower limit of $F$, however, the upper limit is primarily mandated by the prior which we set corresponding to the physical limit of $F=1$. As such, we sample the lower region of the parameter space more completely than the upper region which may lead to biases in our other results. In particular, $F$ has a negative correlation with $H_0$ and hence sampling more points with lower $F$ values corresponds to more points with higher $H_0$ values therefore explaining the increase in $H_0$ from 58$^{+13}_{-9}$\,km\,s$^{-1}$\,Mpc$^{-1}$ to 64$^{+15}_{-13}$\,km\,s$^{-1}$\,Mpc$^{-1}$.

Allowing $F$ to vary also causes a significant steepening in $\alpha$ by more than a factor of two. However, the uncertainties increase by a similar factor and hence the results are still consistent within 1\,$\sigma$.

\end{document}